\documentstyle [aps,float,multicol,prl,psfig]{revtex}
\begin{document}
\title{A scheme for electrical detection of spin resonance signal from a single-electron trap}
\author{I. Martin,$^1$ D. Mozyrsky,$^1$ and H. W. Jiang$^2$}
\address{$^1$Theoretical Division, Los Alamos National Laboratory, Los Alamos, NM 87545,
USA\\
$^2$Department of Physics and Astronomy, University of California at Los Angeles,
Los Angeles, California 90095}
\date{Printed \today}
\maketitle
\begin{abstract}
We study a scheme for electrical detection of the spin resonance (ESR) of a
single electron trapped near a Field Effect Transistor (FET) conduction
channel.  In this scheme, the resonant Rabi oscillations of the trapped
electron spin cause a modification of the average occupancy of a shallow trap,
which can be detected through the change in the FET channel resistivity.  We
show that the dependence of the channel resistivity on the frequency of the rf
field can have either peak or dip at the Larmor frequency of the electron spin
in the trap.
\end{abstract}
\pacs{PACS Numbers: XXXXX}
\begin{multicols}{2}

There has been a lot of interest recently in a few and single electron spin
detection and measurement.  The motivation comes primarily from quantum
computing, where ability to manipulate and to measure single spin is the basis
for several architecture proposals\cite{QC}.  There is also significant
interest in the study of {\em local} electronic environment, e.g. by means of
local electron spin resonance.  Such information would be valuable both for the
conventional semiconductor industry, which has to deal with continuously
decreasing feature sizes, as well as such novel research directions as
spintronics where the utilization of electronic spin degrees of freedom may
lead to conceptually new devices\cite{wolf}.

The main difficulty of a few spin detection and measurement lies in the
inherent weakness of magnetic interaction, making direct measurement of a small
number of spins challenging.  Current state-of-the-art direct detection
techniques, e.g. Magnetic Resonance Force Microscopy\cite{mrfm}, have only
recently achieved the sensitivity of about 100 fully polarized electron spins.
An alternative approach is to {\em convert spin dynamics into charge dynamics}.
Unlike single spin effects, single electron charge signals are much easier to
measure.  For instance, it is well established that the events of capture and
release of electron by a single trap near conducting channel in a field effect
transistor can be measured, either as a Random Telegraph Noise
(RTS)\cite{buhrman,tsai}, or in charge pumping experiments\cite{groes}.  Here
we analyze a setup in which electron spin dynamics under electron spin
resonance conditions is transformed into charge dynamics of the trap occupancy.
This setup is motivated by the recent experiments of M. Xiao and H.W.
Jiang\cite{jiang}, who analyze changes in the statistics of RTS jumps as a
manifestation of the electron spin resonance.

There is a variety of traps that can occur in real systems.  Here we consider
two representative cases schematically shown in Fig. 1 and Fig. 2. They
correspond to the $spinless$ and $spinful$ ``empty'' states of the electron
trap near the FET channel. The conduction channel chemical potential $\mu$ can
be varied with a gate. The single electron levels in the trap are split by the
external magnetic field
$B_0$, $\epsilon_{-1/2} - \epsilon_{1/2} = g\beta B_0$, where
$g$ is the electron g-factor in the trap and $\beta$ is the Bohr magneton.
There is an oscillating magnetic field $B_{\rm rf}(t)$, applied perpendicular
to the field $B_0$ that couples the spin-split single-electron levels. The trap
can accommodate up to two electrons.   The main difference between the two
cases, Fig.~1 and Fig.~2, is that in the presence of near-resonant rf field,
the average electron occupancy of the spinless trap decreases, while for the
spinful trap it increases.   The change in the occupancy of the trap modifies
the FET channel resistivity.  The filling of the trap can cause both increase
and decrease of the channel resistivity, depending on the initial charge state
of the trap.  Hence, there are four distinct combinations of the spin and
charge states of the trap that lead to either enhancement or reduction of the
FET channel resistivity at the resonance (Table~\ref{tab:feat}).

Both spinless and spinful trap cases (Fig.~1 and Fig.~2) can be modelled with
the Anderson Hamiltonian
\begin{eqnarray}
H = \sum_{s} \left(\epsilon_s n_s +{U \over 2} n_s n_{-s}\right) +
\sum_{q,s}\epsilon_{qs}c^\dag_{qs}
c_{qs}&&\nonumber\\
+ \sum_{q,s} T_q \left(c^\dag_{qs} c_s + c^\dag_s c_{qs}\right) + H_{\rm rf}(t)\, . \label{a2}
\end{eqnarray}
In Eq.~(\ref{a2}) $n_s = c^{\dag}_s c_s$, where $c^{\dag}_s(c_s)$ creates
(annihilates) an electron with spin $s=\pm1/2$ in the trap at the level
$\epsilon_s$. The second term in
Eq.~(\ref{a2}) represents Coulomb charging energy. $c^\dag_{qs} (c_{qs})$ are
creation (annihilation) operators for the electrons in the channel. The fourth
term describes tunneling transitions between the trap and the conduction
channel. The tunneling amplitude can be evaluated from the parameters of the
barrier using Bardeen's formula \cite{r02}, or directly from the experiment.
The last term in Eq.~(\ref{a2}),
$H_{\rm rf}(t)$, is the coupling between the spin states in the trap produced by
the {rf} field. If the {rf} frequency is close to that of Zeeman splitting in
the trap, the
$H_{\rm rf}(t)$ can be written in the rotating wave approximation as
\begin{equation}
H_{\rm rf}(t) = \gamma_{rf}\left(c^\dag_{1/2} c_{-1/2} e^{i\omega_{\rm rf} t} + c^\dag_{-1/2}
c_{1/2} e^{-i\omega_{\rm rf} t}\right)\, , \label{a3}
\end{equation}
where $\gamma_{\rm rf} = g \beta B_{\rm rf}/2$, $B_{\rm rf}$ is the amplitude
of the {rf} field. It is assumed that electrons in the conducting channel have
g-factor different from the electronic g-factor in the trap (due to Rashba or
lattice induced spin-orbit coupling). As a result, the influence of the {rf}
field on the electronic states in the channel is ``off-resonance'' and
therefore is neglected in Eq~(\ref{a3}).


As mentioned earlier, even for the simple Anderson model of the trap, there are
four distinct possibilities that correspond to different spin and charge
``empty'' states of the trap.  For concreteness, we consider here two cases:
(1) spinless positive trap, and (2) a neutral trap with spin.  As we will see
both produce a peak in channel resistivity under the ESR resonance conditions.
The other two cases yield a dip.

{\it Positive spinless trap---} The positively charged trap, which essentially
can be a donor impurity, can significantly influence the resistivity of the
conducting channel. Indeed, when the trap is empty it is charged and therefore
acts as a point Coulomb scatterer for the electrons in the channel. On the
other hand, when the trap is filled, it is neutral, thus having relatively
little effect on the channel resistivity \cite{r01}. The average (dc)
resistivity of the channel can be written as
\begin{equation}
\rho = \rho_e \sigma_0 +\rho_f (1 - \sigma_0)\, , \label{a1}
\end{equation}
where $\rho_e$ and $\rho_f$ are channel resistivities with the single electron
trap empty and filled, respectively, and $\sigma_0$ is the probability for the
trap to be empty (throughout the paper, $\sigma$ denotes density matrix, not to
be mistaken for conductivity). We will now demonstrate that due to the coherent
effects in tunneling in and out of the trap induced by the {rf} magnetic field,
the resistivity of the channel
$\rho$ as a function of magnetic field
$B_0$ develops a peak conditioned by $g\beta B_0 = \hbar \omega_{\rm rf}$, where $\omega_{\rm rf}$
is the frequency of the oscillating ESR magnetic field.

The inset in Fig. 1 shows three possible positions of the chemical potential
with respect to the spin levels in the trap. In cases (a) and (c), where
chemical potential lies below and above the two spin levels in the trap,
tunneling into and out of the trap is prohibited. As a result, the electron
occupation number for the trap is independent of the magnetic field at
sufficiently low temperatures (for $k_B T$ smaller than the separation between
the chemical potential and the nearest spin level in the trap), i.e., the trap
is never occupied in case (a) and always occupied in case (c). On the other
hand, if the chemical potential is in between the two spin levels of the trap,
case (b) in the inset, the resonant tunneling into and out of the trap is
possible. That is, an electron with spin ``up'' in the channel can tunnel into
the lowest Zeeman level in the trap, get transmitted into the upper Zeeman
level by the {rf} field and consequently tunnel out of the trap into the
unoccupied levels in the channel. This mechanism is shown schematically in Fig.
1. As a result, electronic occupation number in the trap becomes strongly
modulated by the spin dynamics of the trap and therefore is strongly dependent
on the magnetic field in the vicinity of the spin resonance condition.
\begin{figure}
{\centering{\psfig{figure=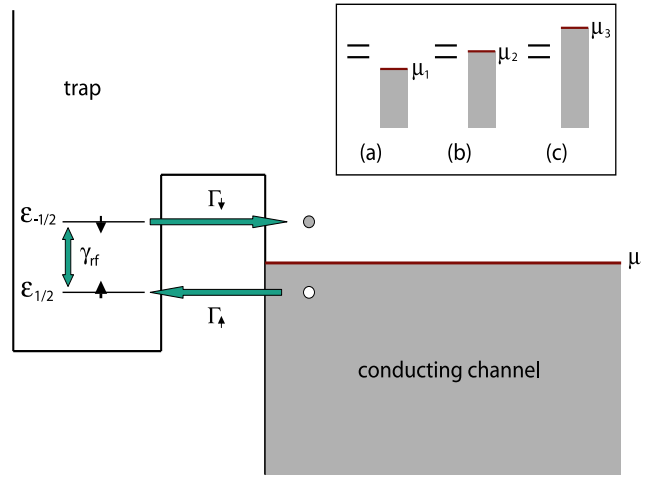,height=7cm,width=8.7cm,angle=90}}}
\vspace{0.5cm} \caption{Model setup for a positively charged spinless trap.
The trap is coupled by the tunneling transitions to a conducting channel. The
trap single-electron energy level is spin-split by the external magnetic field
$B_0$. Chemical potential $\mu$ in the channel (same for both spin species)
is in between the spin-up and spin-down levels in the trap. As a result only
electron with spin ``up'' in the channel can tunnel into the lowest Zeeman
sub-level in the trap, provided it is empty. On the other hand, an electron in
the upper Zeeman sub-level in the trap can tunnel back into the unoccupied
states in the channel. The two levels in the trap are coupled by the
oscillating magnetic field $B_{\rm rf}(t)$. The inset shows energy diagrams
corresponding to three possible positioning of the chemical potential in the
channel relative to the spin levels in the trap.  Only in case (b) ESR-induced
tunneling in and out of the trap is possible.}
\end{figure}

From the Hamiltonian~(\ref{a2},\ref{a3}) one can derive  quantum rate equations
for the electronic occupation numbers in the trap. Similar calculations have
been carried out for a system of two coupled quantum dots in contact with the
Fermi liquid reservoirs \cite{r03}. Omitting technical details we quote the
results of the calculations. We define density matrix for the electronic states
in the trap by introducing density matrix elements $\sigma_{\uparrow\uparrow}$
and
$\sigma_{\downarrow\downarrow}$, which describe probabilities for the electron in the trap to
occupy states with spin ``up'' and ``down'' respectively. We also define the off-diagonal element
$\sigma_{\uparrow\downarrow}$ describing coherent superposition of the ``up'' and ``down'' spin
states in the trap. Together with $\sigma_0$, the probability for the trap to be unoccupied, these
form a set of coupled equations:
\begin{mathletters}
\label{a4}
\begin{eqnarray}
&&{\dot \sigma}_0 = -\Gamma_\uparrow\sigma_0 + \Gamma_\downarrow \sigma_{\downarrow \downarrow}
\,, \label{a4a}\\ &&{\dot \sigma}_{\uparrow\uparrow} = \Gamma_\uparrow\sigma_0 + i(\gamma_{\rm
rf}/\hbar)\left( e^{i\omega_{\rm rf} t}\sigma_{\uparrow\downarrow} - e^{-i\omega_{\rm rf}
t}\sigma_{\downarrow\uparrow}\right)\, , \label{a4b}\\ &&{\dot \sigma}_{\downarrow\downarrow} =
-\Gamma_\downarrow\sigma_{\downarrow\downarrow} - i(\gamma_{\rm rf}/\hbar)\left( e^{i\omega_{\rm
rf} t}\sigma_{\uparrow\downarrow} - e^{-i\omega_{\rm rf}t}\sigma_{\downarrow\uparrow}\right)\, ,
\label{a4c}\\ &&{\dot \sigma}_{\uparrow\downarrow}= -i (E/\hbar) \sigma_{\uparrow\downarrow}
-\Gamma_\downarrow /2 \sigma_{\uparrow\downarrow}\nonumber\\
&&~~~~~~~~~~~~~~~~~~~~~~ + i(\gamma_{\rm rf}/\hbar)e^{-i\omega_{\rm rf} t}
\left(\sigma_{\uparrow\uparrow} - \sigma_{\downarrow\downarrow}\right) \, .\label{a4d}
\end{eqnarray}
\end{mathletters}
In the above equations $E = g\beta B_0$, and $\Gamma_{\uparrow ,\, \downarrow} = (2\pi \eta/\hbar)
T_q^2 (\epsilon = \epsilon_{\uparrow ,\, \downarrow})$, where $\eta$ is the electronic density of
states in the conducting channel (constant for 2DEG) and the tunnel amplitudes $T_q$ are evaluated
at energies of ``up'' and ``down'' spin states in the trap. We assume that the tunnel amplitudes
vary on energy scale much larger than Zeeman splitting in the trap. Therefore in what follows we
put $\Gamma_\uparrow = \Gamma_\downarrow = \Gamma$, where $\Gamma^{-1}$ is the life-time of the
resonant level in the trap. In derivation of Eqs.~(\ref{a4}) we have set $U=\infty$.

Eqs.~(\ref{a4}), though derived from the microscopic Hamiltonian~(\ref{a2}),
have transparent physical meaning. For example, in Eq.~(\ref{a4a}), the rate of
change of $\sigma_0$ is determined by the loss term -- an electron from the
conduction channel can tunnel into the trap with tunnel rate $\Gamma_\uparrow$,
while an electron in the trap can tunnel back into the channel with rate
$\Gamma_\downarrow$ and generate a hole in the trap -- the gain term in the RHS of Eq.~(\ref{a4a}).
The diagonal density matrix elements are coupled with the off-diagonal elements
in Eqs.~(\ref{a4b})-(\ref{a4d}) (the last terms in these equations) due to the
{rf} component of the magnetic field that induces transitions between different
spin states in the trap. These terms always appear in standard Bloch equations
for a spin under magnetic resonance conditions
\cite{r04}.

The rate equations~(\ref{a4}) can be solved for a stationary state, i.e., for
$t\rightarrow\infty$. One finds that $\sigma_0 = \gamma_{\rm rf}^2/[(E-\hbar\omega_{\rm rf})^2 +
(\Gamma\hbar)^2/4 + 3\gamma_{\rm rf}^2]$ and, using Eq.~(\ref{a1}), we obtain
\begin{equation}
\rho (B) = \rho_f  + {\left(\rho_e -\rho_f\right)\gamma_{\rm rf}^2 \over (g\beta
B_0-\hbar\omega_{\rm rf})^2 + (\hbar\Gamma)^2/4 + 3\gamma_{\rm rf}^2} \, , \label{a5}
\end{equation}
Thus, the resistivity of the channel has a resonance when frequency of the {rf}
field matches the Zeeman frequency of the electron in the trap, corresponding
to the condition of {\it single electron spin magnetic resonance}. The width of
the peak is equal to $((\Gamma\hbar)^2/4 + 3\gamma_{\rm rf}^2)^{1/2}$, i.e., it
is determined by both the amplitude of the {rf} field and by the life-time of
the electron in the trap. When
$\gamma_{\rm rf} \gg \Gamma\hbar$, the height of the peak reaches its maximum
$\left(\rho_e -\rho_f\right)/3$ (relative to $\rho_f$, the value of resistivity
away from the resonance).

{\it Neutral trap with spin---} A peak in the FET channel resistivity can be
generated by a neutral trap as well.  Consider a situation shown in Fig. 2.
Here, the ``empty" trap contains an electron that can occupy up or down spin
states (with energies $\epsilon_{1/2}  = -\epsilon_{-1/2}$). In addition, the
trap can be occupied by two electrons with the total energy $U$.  The chemical
potential now lies in the vicinity the two-electron state of the trap. This
corresponds to the case (b) in the inset of Fig. 2.  Similar to the previous
section, the conversion of the ESR spin dynamics into changes of the trap
occupancy is possible only if the absolute difference of the chemical potential
$\mu$ in the channel and the energy level
$U$ in the trap is smaller than half of the Zeeman splitting, i.e.,
$\epsilon_{1/2}<\mu - U < \epsilon_{-1/2}$.
Then, if electron in the trap occupies $\epsilon_{-1/2}$ level, another
electron in the channel with opposite (up) spin can tunnel into the trap, so
that two electrons in the trap occupy state with total energy $U$. Such process
is energetically allowed, provided
$\epsilon_{q,1/2} = U + \epsilon_{1/2}$. Next, an electron in the trap with spin
``down'' can tunnel out of the trap, thus leaving the remaining electron in the
trap in spin ``up'' state. This process is also possible, as long as
$\epsilon_{q,-1/2} = U + \epsilon_{-1/2}$. Thus, the trap effectively relaxes its
spin, creating a particle and a hole in the channel; see Fig. 2. The spin of
the electron in the trap is subsequently flipped by the {rf} field, and so the
process repeats.
\begin{figure}
{\centering{\psfig{figure=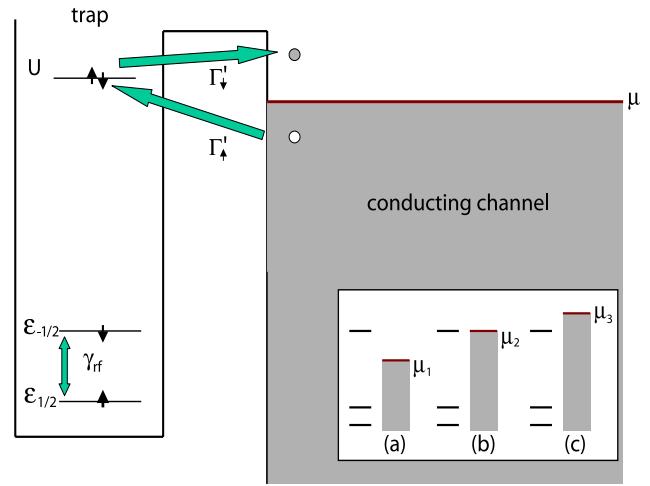,height=7cm,width=8.7cm,angle=90}}}
\vspace{0.5cm} \caption{Model setup for a neutral trap with spin.  The trap is
coupled by the tunneling transitions to a conducting channel. It has a
spin-split single-electron ground state and a two-electron singlet excited
state. Chemical potential $\mu$ in the channel is fixed in the vicinity of the
two-electron excited state; see inset (b). Thus electrons with spin ``up'' in
the channel can tunnel into the excited state $U$ in the trap, provided there
is an electron on
$\epsilon_{-1/2}$ level. An electron with spin down in the trap subsequently
tunnels back into the channel, thus leaving the remaining electron in the trap
in $\epsilon_{1/2}$ state.}
\end{figure}
The resistivity of the channel can now be expressed as
\begin{equation}
\rho = \rho_f \sigma_2 +\rho_e (1 - \sigma_2)\, , \label{a6}
\end{equation}
where $\rho_e$ and $\rho_f$ are resistivities of the trap containing one and
two electrons respectively ($\rho_f > \rho_e$), and $\sigma_2$ is the
probability for the trap to be occupied by two electrons. Again, from the
Hamiltonian~(\ref{a2}) we derive the rate equations for the trap
\begin{mathletters}
\label{a7}
\begin{eqnarray}
&&{\dot \sigma}_2 = -\Gamma_\downarrow^\prime\sigma_2 + \Gamma_\uparrow^\prime\sigma_{\downarrow
\downarrow} \,, \label{a7a}\\ &&{\dot \sigma}_{\uparrow\uparrow} = \Gamma_\downarrow^\prime\sigma_2
+ i(\gamma_{\rm rf}/\hbar)\left( e^{i\omega_{\rm rf} t}\sigma_{\uparrow\downarrow} -
e^{-i\omega_{\rm rf} t}\sigma_{\downarrow\uparrow}\right)\, , \label{a7b}\\ &&{\dot
\sigma}_{\downarrow\downarrow} = -\Gamma_\uparrow^\prime\sigma_{\downarrow\downarrow} -
i(\gamma_{\rm rf}/\hbar)\left( e^{i\omega_{\rm rf} t}\sigma_{\uparrow\downarrow} - e^{-i\omega_{\rm
rf}t}\sigma_{\downarrow\uparrow}\right)\, , \label{a7c}\\ &&{\dot \sigma}_{\uparrow\downarrow}= -i
(E/\hbar) \sigma_{\uparrow\downarrow}
-\Gamma_\uparrow^\prime /2 \sigma_{\uparrow\downarrow}\nonumber\\
&&~~~~~~~~~~~~~~~~~~~~~~ + i(\gamma_{\rm rf}/\hbar)e^{-i\omega_{\rm rf} t}
\left(\sigma_{\uparrow\uparrow} - \sigma_{\downarrow\downarrow}\right) \, .\label{a7d}
\end{eqnarray}
\end{mathletters}
In Eqs.~(\ref{a7}) $\sigma$'s are defined above, while the tunnel rates $\Gamma_{\uparrow ,\,
\downarrow}^\prime = (2\pi \eta/\hbar) T_q^2 (\epsilon = \epsilon_{\uparrow ,\, \downarrow}+U)$.
Assuming $\Gamma_\uparrow^\prime = \Gamma_\downarrow^\prime = \Gamma^\prime$,
Eqs.~(\ref{a7}) can be solved for a stationary state. Then, substituting thus
obtained $\sigma_2$ into the Eq.~(\ref{a6}), we find an expression for the
resistivity similar to Eq.~(\ref{a5}), but with
$\rho_e$ and $\rho_f$ interchanged and $\Gamma$ replaced by $\Gamma^\prime$. Thus, the resistivity
of the channel as a function of magnetic field $B$ again has a peak centered at
$\omega_{\rm rf}$.

{\it Effects of spin relaxation---} Effects of environment, such as phonons,
nuclear spins, etc., can be taken into account by introducing additional
relaxation rate, $1/T_2^\prime$ in the equations for the off-diagonal elements
of the spin (here we neglect the longitudinal spin relaxation, i.e., $T_1
\gg T_2^\prime,\Gamma$).  Note, that the noise spectra that induce $1/T_2^\prime$ must be
taken at the Rabi ($\gamma_{\rm rf}$), and not zero, frequency\cite{r04}.
Thus, replacing rates $\Gamma_\downarrow/2$ and $\Gamma_\uparrow^\prime/2$  in
Eqs.~(\ref{a4d},~\ref{a7d}) by $D = \Gamma_\downarrow /2 + 1/T_2^\prime$ and
$D^\prime = \Gamma_\uparrow^\prime /2 + 1/T_2^\prime$, and repeating the above
algebra, we obtain the resistivity of the channel in the presence of externally
induced spin dephasing
\begin{equation}
\rho (B) = \rho_f  + {2\left(\rho_e -\rho_f\right)\gamma_{\rm rf}^2 \over (g\beta
B_0-\hbar\omega_{\rm rf})^2(\Gamma/D) + \hbar^2\Gamma D + 6\gamma_{\rm rf}^2} \, , \label{a8}
\end{equation}
for the positively charged trap, while $\rho (B)$ for the neutral trap is obtained by replacing
$\Gamma$ and $D$ in the above equation by $\Gamma^\prime$ and $D^\prime$. Eq.~(\ref{a8}) shows
that the height of the peak in the resistivity decreases due to the suppression
of the spin coherence by the externally induced dephasing rate $1/T_2^\prime$,
and the peak width becomes $\sqrt{D^2 + 6(\gamma_{\rm rf}/\hbar)^2 D/\Gamma}$.

{\it Rabi oscillations of a single electron trap ---} The electrical detection
of Rabi oscillations from a single trap can also be performed  under the
conditions described above.  Rabi oscillations correspond to the coherent
weight transfer between the states of a two level system.  Therefore, in the
case of the spinless trap at the ESR resonance, the probability to tunnel out
of the trap between times $t$ and $t + dt$ after the trapping, $dP(t) = \Gamma
e^{-\Gamma t} dt (1 - \cos(2\gamma_{\rm rf} t))$, is periodically modulated at
the Rabi frequency (here we assumed that there is no extrinsic dephasing,
$1/T_2^\prime = 0$). Since the current between the trapping and escape events
remains constant, the current power spectrum can be easily calculated as
\begin{equation}
\langle |I_\omega|^2 \rangle \propto \frac{1}{\omega^2} \left( 2 +
\frac{\Gamma^2}{\Gamma^2 + (\omega - 2\gamma_{\rm rf})^2}\right),
\end{equation}
for $\omega \gg \Gamma$. In case the externally induced dephasing dominates, in
the above expression $\Gamma$ should be replaced by $1/T_2^\prime$.

In summary, we proposed a mechanism for electrical detection of electron spin
resonance from a single electron trapped near the conducting channel of a field
effect transistor.  The dc effect is based on the modification of the {\em
average} trap occupancy caused by the resonant excitation of the trap.
Depending on the charge and spin propertied of the empty trap, at the
resonance, the channel resistivity can both increase and decrease, an effect
that can be used for trap diagnostics.  We also propose a scheme for the Rabi
oscillation detection in the same system.

\begin{table}
  \centering
  \caption{Feature in the FET channel resistivity as a function of
 the empty state of the electron trap}\label{tab:feat}
\begin{tabular}{l c c}
           & neutral or negative & positive \\\cline{1-3}
  spinless  & dip & peak \\
  spinful  & peak & dip \\
\end{tabular}
\end{table}

{\it Acknowledgements---} This work was supported by the U.S. DOE and DARPA
SPINs program.  D.M. was supported in part by the NSF grants DMR-0121146 and
ECS-0102500.

\end{multicols}

\begin{references}
\bibitem{r01} If electrons in the channel are polarized, the resistivity can be
also weakly modulated by the relative orientation of the electronic spin in the
trap and of the conduction electron spin in the channel due to their exchange
interaction.
\end{references}
\end{document}